\documentclass[useAMS, usenatbib, a4paper]{mnras}
\usepackage{graphicx}
\usepackage{microtype}
\usepackage{xcolor}
\usepackage{fixltx2e}
\usepackage{booktabs}
\usepackage{siunitx}
\usepackage{color}
\usepackage{enumerate}
\usepackage{pdflscape}
\usepackage{rotating}
\usepackage{xr-hyper}
\usepackage{hyperref}

\usepackage[T1]{fontenc} 
\usepackage[utf8]{inputenc}

\usepackage{newtxtext}
\usepackage[varvw,smallerops]{newtxmath}

\usepackage{chemgreek}
\activatechemgreekmapping{newtx}
\usepackage{listings}

\hypersetup{colorlinks=True, linkcolor=blue!50!black, citecolor=black,
  urlcolor=blue!50!black}

\usepackage{etoolbox}
\robustify\bfseries
\robustify\itshape

\makeatletter
\patchcmd\@combinedblfloats{\box\@outputbox}{\unvbox\@outputbox}{}{%
  \errmessage{\noexpand\@combinedblfloats could not be patched}%
}%
\makeatother

\usepackage{bm}

\usepackage{aastex-compat}

\title
{Bow shocks, bow waves, and dust waves. I. Strong coupling limit}

\newcommand\AddressCRyA{Instituto de Radioastronom\'{\i}a y Astrof\'{\i}sica,
  Universidad Nacional Aut\'onoma de M\'exico, Apartado Postal 3-72,
  58090 Morelia, Michoac\'an, M\'exico}
\author[Henney \& Arthur]{
  William J. Henney\thanks{w.henney@irya.unam.mx}
  \& S. J. Arthur\\
  \AddressCRyA
}

\date{Accepted XXX. Received YYY; in original form ZZZ}

\pubyear{2019}

\newcommand{\grain}{\ensuremath{_{\text{d}}}}

\newcommand{\wind}{\ensuremath{_{\text{w}}}}

\newcommand\rad{\ensuremath{_{\text{rad}}}}

\newcommand\sound{\ensuremath{c_{\text{s}}}}

\newcommand\starstar{\ensuremath{_{**}}}

\newcommand\alphaB{\ensuremath{\alpha_{\text{B}}}}
\newcommand\shell{\ensuremath{_{\text{sh}}}}
\newcommand\M{\ensuremath{\mathcal{M}}}
\newcommand\hii{\ion{H}{ii}}

\defcitealias{Tarango-Yong:2018a}{Paper~I}

\begin{document}
\label{firstpage}
\pagerange{\pageref{firstpage}--\pageref{lastpage}}
\maketitle
\begin{abstract}
  Dust waves and bow waves result from the action of a star's
  radiation pressure on a stream of dusty plasma that flows past it.
  They are an alternative mechanism to hydrodynamic bow shocks for
  explaining the curved arcs of infrared emission seen around some
  stars.  When gas and grains are perfectly coupled, for a broad class
  of stellar parameters, wind-supported bow shocks predominate when
  the ambient density is below \SI{100}{cm^{-3}}.  At higher densities
  radiation-supported bow shells can form, tending to be optically
  thin bow waves around B~stars, or optically thick bow shocks around
  early O~stars.  For OB stars with particularly weak stellar winds,
  radiation-supported bow shells become more prevalent.
\end{abstract}


\section{Introduction}
\label{sec:introduction}

\begin{figure*}
  \centering
  \includegraphics[width=0.8\linewidth]{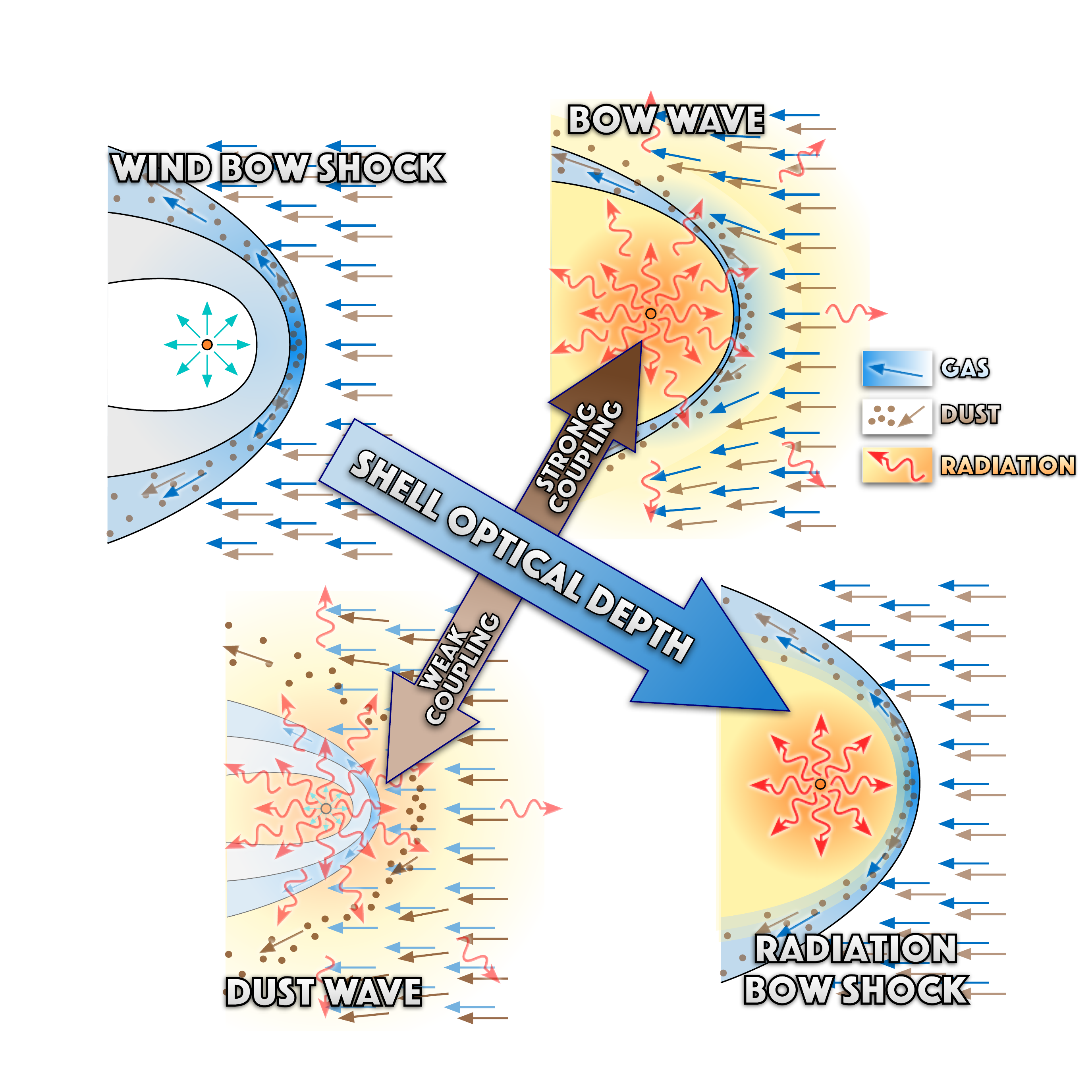}
  \caption{Regimes in the supersonic interaction of a luminous star
    with its environment.  When radiation effects are unimportant, we
    have the standard (magneto-)hydrodynamical wind-supported bow
    shock (upper left), where the ram pressure of the stellar wind
    balances the ram pressure of the oncoming stream.  As the optical
    depth of the shocked shell increases, the stellar radiation
    momentum adds to the wind ram pressure to help support the bow
    shell.  If the shell is completely opaque to stellar radiation, we
    have a radiation-supported bow shock (lower right), where it is
    the stellar radiation pressure that balances the ram pressure of
    the external stream.  For intermediate optical depths, we have two
    cases depending on the strength of coupling between grains and
    gas.  If the coupling is strong, then we have a
    radiation-supported bow wave (upper right), where the plasma
    stream as a whole is gradually radiatively decelerated as it
    approaches the star.  If the coupling is weak, then the radiation
    momentum is felt only by the dust, which decouples from the gas to
    form a dust wave, with the gas stream continuing inward to form a
    wind-supported bow shock closer to the star.  Note that the
    gas--grain coupling is both direct (via collisions) and indirect
    (via the magnetic field) (see Paper~II).}
  \label{fig:3-types-bow}
\end{figure*}

Curved emission arcs around stars \citep[e.g.,][]{Gull:1979a} are
often interpreted as \textit{bow shocks}, due to a supersonic
hydrodynamic interaction between the star's wind and an external
stream. This stream may be due to the star's own motion or to an
independent flow, such as an \hii{} region in the champagne phase
\citep{Tenorio-Tagle:1979a}, or another star's wind
\citep{Canto:1996}. However, an alternative interpretation in some
cases may be a radiation-pressure driven bow wave, as first proposed
by \citet[\S\textsc{vi}]{van-Buren:1988a}.  In this scenario (see
Fig.~\ref{fig:3-types-bow}), photons emitted by the star are absorbed
by dust grains in the incoming stream, with the resultant momentum
transfer being sufficient to decelerate and deflect the grains within
a certain distance from the star, forming a dust-free, bow-shaped
cavity with an enhanced dust density at its edge.

Two regimes are possible, depending on the strength of coupling
between the gas (or plasma) and the dust.  In the strong-coupling
regime, gas--grain drag decelerates the gas along with the dust.  If
the stream is optically thin to the star's ultraviolet radiation, then
the deceleration occurs gradually over a range of radii, forming a
relatively thick shell.  On the other hand, if the stream is optically
thick, then a shocked gas shell forms in a similar fashion to the
wind-driven bow shock case, except internally supported by trapped
radiation instead of shocked stellar wind.  In the weak-coupling
regime, the gas stream is relatively unaffected and the dust
temporarily decouples to form a dust-only shell.  This second case has
recently been studied in detail in the context of the interaction of
late O-type stars (some of which have very weak stellar winds) with
dusty photoevaporation flows inside \hii{} regions
\citep{Ochsendorf:2014a, Ochsendorf:2014b, Ochsendorf:2015a}.  We
follow the nomenclature proposed by \citet{Ochsendorf:2014b}, in which
\textit{dust wave} refers to the weak coupling case and \textit{bow
  wave} to the strong coupling case.  More complex, hybrid scenarios
are also possible, such as that studied by \citet{van-Marle:2011a},
where a hydrodynamic bow shock forms, but the larger dust grains that
accompany the stellar wind pass right through the shocked gas shell,
and form their own dust wave at a larger radius.

This is the first in a series of papers where we develop simple
physical models to show in detail when and how these different
interaction regimes apply when varying the parameters of the star, the
dust grains, and the ambient stream.  We concentrate primarily on the
case of luminous early type stars, where dust is present only in the
ambient stream, and not in the stellar wind.  In this first paper, we
consider the case where the grains are perfectly coupled to the gas
via collisions.  The following two papers consider the decoupling of
grains and gas in a sufficiently strong radiation field
\citep[Paper~II]{Henney:2019b}, and how observations can distinguish
between different classes of bow shell
\citep[Paper~III]{Henney:2019c}.
The paper is organized as follows.
In \S~\ref{sec:strong-gas-grain} we propose a simple model for stellar
bow shells and investigate the relative importance of wind and radiation in
providing internal support for the bow shell as a function of the
density and velocity of the ambient stream, and for different types of
star.  In \S~\ref{sec:phys-state-shock} we calculate the physical
state of the bow shell, considering under what circumstances it can
trap within itself the star's ionization front and how efficient
radiative cooling will be.
In \S~\ref{sec:discussion} we briefly discuss the application
of our models to observed bow shells.
In \S~\ref{sec:summary} we summarise our findings. 
%



\section{A simple model for stellar bow shells}
\label{sec:strong-gas-grain}

\begin{figure*}
  \includegraphics[width=\linewidth]{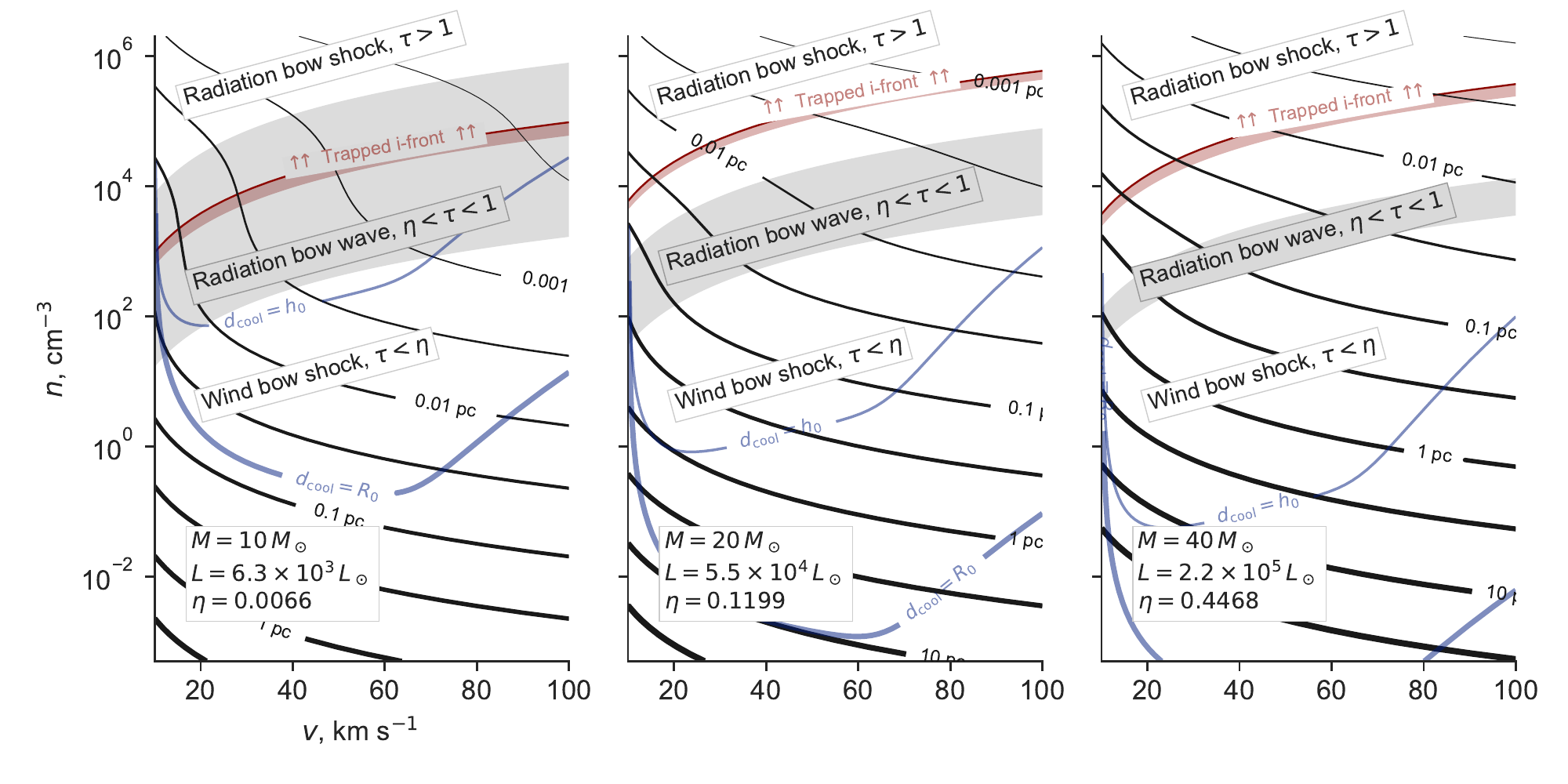}
  \caption{Bow shell regimes in parameter space (\(v, n\)) of the
    external stream for main-sequence OB stars of different masses:
    (a)~\SI{10}{M_\odot}, (b)~\SI{20}{M_\odot}, (c)~\SI{40}{M_\odot}.  In all
    cases, \(\kappa = \SI{600}{cm^2.g^{-1}}\) and efficient gas-grain
    coupling is assumed. Solid black lines of varying thickness show
    the bow shell size (star-apex separation, \(R_0\)), while gray
    shading shows the radiation bow wave regime, with lower border
    \(\tau = \eta\wind\) and upper border \(\tau = 1\), where
    \(\tau = 2 \kappa \rho R_0\) is the optical depth through the bow.  For bow
    shells above the red solid line, the ionization front is trapped
    inside the bow shell (\S~\ref{sec:trapp-ioniz-front}).  Blue lines
    delineate different cooling regimes
    (\S~\ref{sec:radi-cool-lengths}).  Above the thin blue line
    (\(d_{\text{cool}} = h_0\)), the bow shock radiates efficiently,
    forming a thin shocked shell.  Below the thick blue line
    (\(d_{\text{cool}} = R_0\)), the bow shock is essentially
    non-radiative.}
  \label{fig:zones-v-n-plane}
\end{figure*}

We consider the canonical case of a bow shell around a star of
bolometric luminosity, \(L\), with a radiatively driven wind, which is
immersed in an external stream of gas and dust with density, \(\rho\),
and velocity, \(v\).  The size and shape of the bow shell is
determined by a generalized balance of pressure (or, equivalently,
momentum) between internal and external sources.  We assume that the
stream is supersonic and super-alfvenic, so that the external pressure
is dominated by the ram pressure, \(\rho v^2\), and that dust grains and
gas are perfectly coupled by collisions (the breakdown of this
assumption is the topic of Paper~II).

Although dust grains typically constitute only a small fraction
\(Z\grain \sim 0.01\) of the mass of the external stream, they
nevertheless dominate the broad-band opacity at FUV, optical and IR
wavelengths if they are present. They also dominate at EUV wavelengths
(\(\lambda < \SI{912}{\angstrom}\)), if the hydrogen neutral fraction is
less than \(\approx 0.001\).  The strong coupling assumption means that all
the radiative forces applied to the dust grains are directly felt by
the gas also.

\subsection{Bow shells supported by radiation and wind}
\label{sec:three-bow-regimes}

The internal pressure is the sum of wind ram pressure and the
effective radiation pressure that acts on the bow shell.  The
radiative momentum loss rate of the star is \(L/c\) and the wind
momentum loss rate can be expressed as
\begin{equation}
  \label{eq:wind-efficiency}
  \dot{M} V = \eta\wind L / c \ , 
\end{equation}
where \(\eta\wind\) is the momentum efficiency of the wind, which is
\(< 1\) in all cases except WR stars \citep{Lamers:1999b}. If the
optical depth at UV/optical wavelengths is sufficiently large, then
all of the stellar radiative momentum, emitted with rate \(L/c\), is
trapped by the bow shell.  At the infrared wavelengths where the
absorbed radiation is re-emitted, the grain opacity is much lower, so
the bow shell is optically thin to this radiation.  Combined with the fact
that the UV single-scattering albedo is only \(\approx 0.5\), this means
that the single-scattering limit is approximately valid, and we adopt
it here.

We define a fiducial bow shock radius \(R_*\) in the optically thick,
radiation-only limit by balancing stellar radiation pressure against
stream ram pressure at the bow apex along the
symmetry axis from the star:
\begin{equation}
  \label{eq:rad-press-balance-thick}
  \frac{L}{4 \pi c R_*^2} = \rho v^2 \ ,
\end{equation}
which yields 
\begin{equation}
  \label{eq:Rstar}
  R_* = \left(\frac{L}{4\pi c \rho v^2}\right)^{1/2} \ .
\end{equation}

We now consider the opposite, optically thin limit.  If the total
opacity (gas plus dust) per total mass (gas plus dust) is \(\kappa\) (with
units of \si{cm^2.g^{-1}}), then the radiative acceleration is
\begin{equation}
  \label{eq:rad-accel}
  a_{\text{rad}} = \frac{\kappa L}{4 \pi c R^2} \ .
\end{equation}
Therefore, an incoming stream with initial velocity, \(v_\infty\), can be
brought to rest by radiation alone at a distance \(R\starstar\) where
\begin{equation}
  \label{eq:rad-poten}
  \int_{R\starstar}^\infty a_{\text{rad}} \, dr = \tfrac12 v_\infty^2 \ , 
\end{equation}
yielding
\begin{equation}
  \label{eq:rad:R0}
  R\starstar = \frac{\kappa L}{2\pi c v_\infty^2} \ .
\end{equation}
On the other hand, we can also argue as in the optically thick case
above by approximating the bow shell as a surface, and balancing
stellar radiation pressure against the ram pressure of the incoming
stream.  The important difference when the shell is not optically
thick is that only a fraction \(1 - e^{-\tau}\) of the radiative momentum
is absorbed by the bow, so that
equation~\eqref{eq:rad-press-balance-thick} is replaced with
\begin{equation}
  \label{eq:rad-press-balance-tau}
  \frac{L (1 - e^{-\tau})}{4 \pi c R_0^2} = \rho v^2 \ .
\end{equation}
In the optically thin limit, \(1 - e^{-\tau} \approx \tau\), so these two
descriptions can be seen to agree (\(R_0 \to R\starstar\)) so long as
\begin{equation}
  \label{eq:tau-thin}
  \tau = 2 \kappa \rho R_0 \ ,
\end{equation}
which we will assume to hold generally.

We define a fiducial optical depth,
\begin{equation}
  \label{eq:tau-star}
  \tau_* = \rho \kappa R_* \ .
\end{equation}
With these fiducial values established, we return to the combined
radiation plus wind scenario.  Adding the stellar wind ram pressure
term from equation~\eqref{eq:wind-efficiency} allows us to write the
general bow shell radius in terms of the fiducial radius as
\begin{equation}
  \label{eq:R0-definition}
  R_0 = x R_* \ ,
\end{equation}
where \(x\) is the solution of
\begin{equation}
  \label{eq:rad-full-x}
  x^2 - \bigl(1 - e^{-2 \tau_* x} \bigr) - \eta\wind = 0 \ .
\end{equation}
Since this is a transcendental equation, \(x\) must be found
numerically, but we can write explicit expressions for three limiting
cases:
\begin{equation}
  \label{eq:x-cases}
  x \approx
  \begin{cases}
    \text{if \(\tau_* \gg 1\):} & (1 + \eta\wind)^{1/2}  \\
    \text{if \(\tau_*^2 \ll 1\):} & \tau_* + \bigl( \tau_*^2 + \eta\wind \bigr)^{1/2} \approx
    \begin{cases}
      \text{if \(\tau_*^2 \gg \eta\wind\):} & 2 \tau_*  \\
      \text{if \(\tau_*^2 \ll \eta\wind\):} & \eta\wind^{1/2} 
    \end{cases}
  \end{cases}
\end{equation}
The first case, \(x \approx (1 + \eta\wind)^{1/2}\), corresponds to a
\textit{radiation bow shock} (RBS); the second case,
\(x \approx 2 \tau_* \), corresponds to a \textit{radiation bow wave} (RBW);
and the third case, \(x \approx \eta\wind^{1/2}\), corresponds to a
\textit{wind bow shock} (WBS).  The two bow shock cases are similar in
that the external stream is oblivious to the presence of the star
until it suddenly hits the bow shock shell, differing only in whether
it is radiation or wind that is providing the internal pressure.  In
the intermediate bow wave case, on the other hand, the external stream
is gradually decelerated by absorption of photons as it approaches the
bow.  We remark that a shock can still form in this case, but shocked
material constitutes only a fraction of the total column density of
the shell, as we will show in \S~\ref{sec:axial-structure-bow}.

\subsection{Dependence on stellar type}
\label{sec:depend-stell-type}

\begin{table*}
  \centering
  \caption{Stellar parameters for example stars}
  \label{tab:stars}
  \begin{tabular}{l S S S S S S S S S l}
    \toprule
    & {\(M / \si{M_\odot}\)} & {\(L_4\)}
    & {\(\dot{M}_{-7}\)} & {\(V_3\)} & {\( \eta\wind \)}
    & {Sp.~Type} 
    & {\(T_{\text{eff}} / \si{kK}\)} & {\(\lambda_{\text{eff}}\) / \si{\um}}
    & {\(S_{49}\)} & Figures 
    \\
    \midrule
    & 10 & 0.63 & 0.0034 & 2.47 & 0.0066 & {B1.5\,V} & 25.2 & 0.115 & 0.00013
                   & \ref{fig:zones-v-n-plane}a \\
    Main-sequence OB stars
    & 20 & 5.45 & 0.492 & 2.66 & 0.1199 & {O9\,V} & 33.9 & 0.086 & 0.16
                   & \ref{fig:zones-v-n-plane}b, \ref{fig:O-weak-wind} \\
    & 40 & 22.2 & 5.1 & 3.31 & 0.4468 & {O5\,V} & 42.5 & 0.068 & 1.41
                   & \ref{fig:zones-v-n-plane}c\\[\smallskipamount]
    Blue supergiant star
    & 33 & 30.2 & 20.2 & 0.93 & 0.3079 & {B0.7\,Ia} & 23.5 & 0.123 & 0.016
                   & \ref{fig:B-supergiant} \\[\smallskipamount]
    Red supergiant star
    & 20 & 15.6 & 100 & 0.015 & 0.0476 & {M1\,Ia} & 3.6 & 0.805 & 0
                   & \ref{fig:M-supergiant} \\ 
    \bottomrule
  \end{tabular}
\end{table*}

We now consider the application to bow shocks around main sequence OB
stars, as well as cool and hot supergiants, expressing stellar and
ambient parameters in terms of typical values as follows:
\begin{align*}
  \label{eq:stellar-parameters}
  \dot{M}_{-7} &= \dot{M} / \bigl(\SI{e-7}{M_\odot.yr^{-1}}\bigr) \\
  V_3 &= V / \bigl(\SI{1000}{km.s^{-1}}\bigr) \\
  L_4 &= L / \bigl(\SI{e4}{L_\odot}\bigr) \\
  v_{10} &= v_\infty / \bigl( \SI{10}{km.s^{-1}} \bigr) \\
  n &= (\rho / \bar{m}) / \bigl( \SI{1}{cm^{-3}} \bigr) \\
  \kappa_{600} &= \kappa / \bigl( \SI{600}{cm^2.g^{-1}} \bigr) \ ,
\end{align*}
where \(\bar{m}\) is the mean mass per hydrogen nucleon
(\(\bar{m} \approx 1.3 m_{\text{p}} \approx \SI{2.17e-24}{g}\) for solar
abundances).  Note that \(\kappa = \SI{600}{cm^2.g^{-1}}\) corresponds to a
cross section of \(\approx \SI{e-21}{cm^2}\) per hydrogen nucleon, which is
typical for interstellar medium dust \citep{Bertoldi:1996a} at far
ultraviolet wavelengths, where OB stars emit most of their radiation.
In terms of these parameters, we can express the stellar wind momentum
efficiency as
\begin{equation}
  \label{eq:wind-eta-typical}
  \eta\wind = \num{0.495} \,\dot{M}_{-7} \,V_3  \,L_4^{-1}
\end{equation}
and the fiducial radius and optical depth as
\begin{align}
  \label{eq:Rstar-typical}
  R_* / \si{pc} &= \num{2.21} \, (L_4 / n)^{1/2} \,v_{10}^{-1} \\
  \label{eq:taustar-typical}
  \tau_* &= \num{0.0089} \,\kappa_{600} \, (L_4 \,n)^{1/2} \,v_{10}^{-1} \ .
\end{align}
In Figure~\ref{fig:zones-v-n-plane}, we show results for the bow size
(apex distance, \(R_0\)) as a function of the density, \(n\), and
relative velocity, \(v_\infty\), of the external stream, with each panel
corresponding to a particular star, with parameters as shown in
Table~\ref{tab:stars}.  To facilitate comparison with previous work,
we choose stellar parameters similar to those used in the
hydrodynamical simulations of \citet{Meyer:2014b, Meyer:2016a,
  Meyer:2017a}, based on stellar evolution tracks for stars of
\SIlist{10;20;40}{M_\odot} \citep{Brott:2011a} and theoretical wind
prescriptions \citep{de-Jager:1988a, Vink:2000a}.  Although the
stellar parameters do evolve with time, they change relatively little
during the main-sequence lifetime of several million years.\footnote{%
  \label{fn:meyer-velocities-too-low}
  Note that we have recalculated the stellar wind terminal velocities,
  since the values given in the \citeauthor{Meyer:2014b} papers are
  troublingly low.  We have used the prescription
  \(V = 2.6 V_{\text{esc}}\), where
  \(V_{\text{esc}} = \left( 2 G M (1 - \Gamma_e)/ R \right)^{1/2}\) is the
  photospheric escape velocity, which is appropriate for strong
  line-driven winds with \(T_{\text{eff}} > \SI{21 000}{K}\)
  \citep{Lamers:1995a}.  We find velocities of
  \SIrange{2500}{3300}{km.s^{-1}}, which are consistent with
  observations and theory \citep{Vink:2000a}, but at least two times
  higher than those cited by \citet{Meyer:2014b}. } %
The three examples are an early B~star (\SI{10}{M_\odot}), a late O~star
(\SI{20}{M_\odot}), and an early O~star (\SI{40}{M_\odot}), which cover the
range of luminosities and wind strengths expected from bow-producing
hot main sequence stars.  The luminosity is a steep function of
stellar mass (\(L \sim M^{2.5}\)) and the wind mass-loss rate is a steep
function of luminosity (\(\dot{M} \sim L^{2.2}\)), which means that the
wind momentum efficiency is also a steep function of mass
(\(\eta\wind \sim M^3\)), approaching unity for early O~stars, but falling
to less than 1\% for B~stars.

\begin{figure}
  \includegraphics[width=\linewidth]{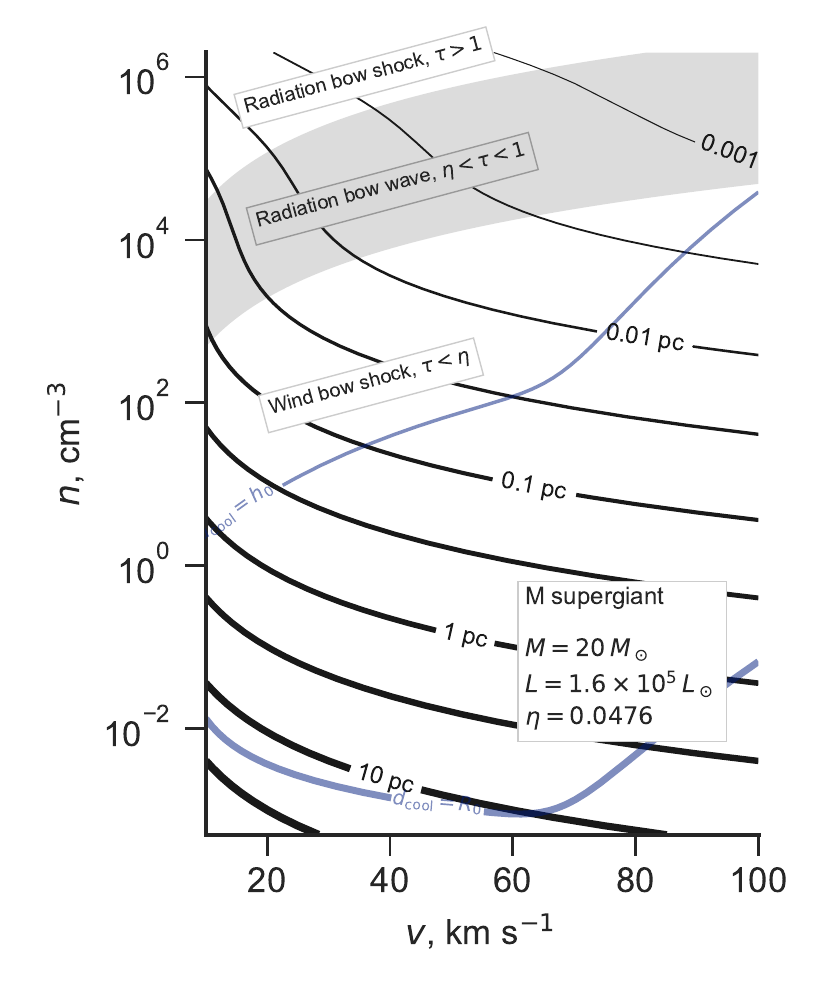}
  \caption{As Fig.~\ref{fig:zones-v-n-plane}, but for a cool M-type
    supergiant instead of hot main sequence stars.  A smaller dust
    opacity is used, \(\kappa = \SI{60}{cm^2.g^{-1}}\), because of the
    reduced extinction efficiency at the optical/infrared wavelengths
    emitted by this star.}
  \label{fig:M-supergiant}
\end{figure}

It can be seen from Figure~\ref{fig:zones-v-n-plane} that the onset of
the radiation bow wave regime is very similar for the three
main-sequence stars, occurring at
\(n > \text{\numrange{20}{40}} \, v_{10}^2\).  An important
difference, however, is that for the \SI{40}{M_\odot} star, which has a
powerful wind, the radiation bow wave regime only occurs for a very
narrow range of densities, whereas for the \SI{10}{M_\odot} star, with a
much weaker wind, the regime is much broader, extending to
\(n < \num{e4} \, v_{10}^2\).  Another difference is the size scale of
the bow shells in this regime, which is
\(R_0 = \text{\SIrange{0.001}{0.003}{pc}}\) for the \SI{10}{M_\odot} star
if \(v_\infty = \SI{40}{km.s^{-1}}\), but \(R_0 \approx \SI{0.1}{pc}\) for the
\SI{40}{M_\odot} star, assuming the same inflow velocity.

Figure~\ref{fig:M-supergiant} shows results for a cool M-type
super-giant star with stellar parameters inspired by Betelgeuse
(\chemalpha~Orionis), as listed in Table~\ref{tab:stars}.  Unlike the
UV-dominated spectrum of the hot stars, this star emits predominantly
in the near-infrared, where the dust extinction efficiency is lower \citep{Weingartner:2001a},
so we adopt a lower opacity of \SI{60}{cm^2.g^{-1}}.  This has the
effect of shifting the radiation bow wave regime to higher densities:
\(n = \text{\numrange{1000}{30 000}}\, v_{10}^2\) in this case.

\subsection{Effects of stellar gravity}
\label{sec:effects-gravity}

In principle, gravitational attraction from the star, of mass \(M\),
will partially counteract the radiative acceleration.  This can be
accounted for by replacing \(L\) with an effective luminosity
\newcommand\Edd{\ensuremath{_{\text{E}}}}
\begin{equation}
  \label{eq:effective-luminosity}
  L_{\text{eff}} = L \bigl(1 - \Gamma\Edd^{\,-1}\bigr) \ ,
\end{equation}
in which \(\Gamma\Edd\) is the Eddington factor:
\begin{equation}
  \label{eq:eddington-factor}
  \Gamma\Edd = \frac{\kappa L}{4\pi c G M} = 458.5 \, \frac{\kappa_{600} L_4}{ M } \ ,
\end{equation}
where, in the last expression, \(M\) is measured in solar masses.  For
the stars in Table~\ref{tab:stars}, we find
\(\Gamma\Edd \approx \text{\numrange{30}{400}}\), so gravity can be safely
ignored.  When the optical depth of the bow shell is very large,
\(\tau > \ln\Gamma\Edd \sim 5\), gravity will exceed the radiation force in the
outer parts of the shell (see \citealt{Rodriguez-Ramirez:2016b}), but
it is generally too weak to affect the shell structure even in such a
case, see \S~\ref{sec:axial-structure-bow} below.

\section{Physical state of the bow shell}
\label{sec:phys-state-shock}

There are two shocked zones in the bow: an inner zone of shocked
stellar wind, and an outer zone of decelerated ambient stream. For OB
stars, dust grains are present only in the ambient stream, whereas for
cool stars they will also be present in the stellar wind.  In the
remainder of this paper, we will concentrate on the OB star case,
where it is the outer zone that is most important observationally.
Bow shells are typically detected via their infrared radiation (absorption
and re-emission by dust of stellar radiation), or by emission lines
such as the hydrogen H\(\alpha\) line.


\subsection{Ionization state}
\label{sec:trapp-ioniz-front}

\begin{figure}
  \includegraphics[width=\linewidth]{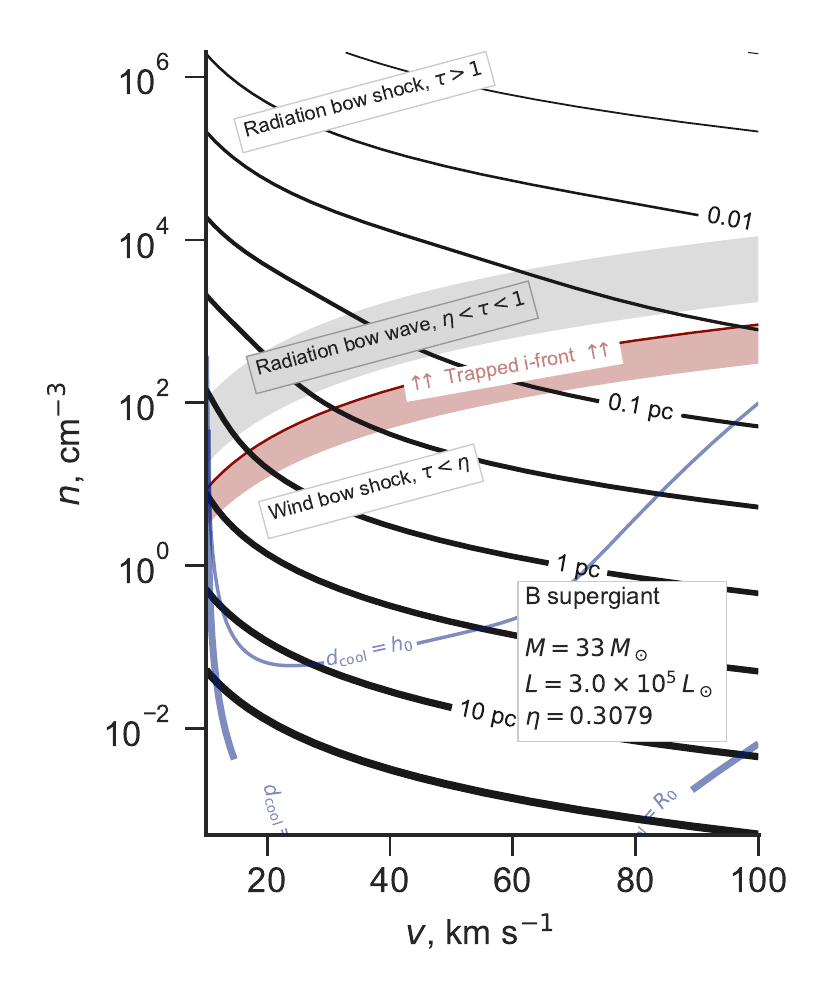}
  \caption{As Fig.~\ref{fig:zones-v-n-plane}, but for an evolved
    B-type supergiant instead of main sequence stars.  This is similar
    to the early O MS star of Fig.~\ref{fig:zones-v-n-plane}\textit{c}
    in many respects, except for the trapping of the ionization front,
    which occurs for much lower outer stream densities.}
  \label{fig:B-supergiant}
\end{figure}

In this section we calculate whether the star is capable of
photoionizing the entire bow shock shell, or whether the ionization
front will be trapped within it.  Using the on-the-spot approximation
for the diffuse fields \citep{Osterbrock:2006a}, the number of
hydrogen recombinations per unit time per unit area in a fully ionized
shell of density \(n\shell\) and thickness \(h\shell\) is
\begin{equation}
  \label{eq:shell-recombination-rate}
  \mathcal{R} = \alphaB n\shell^2 h\shell \ ,
\end{equation}
where \(\alphaB = \num{2.6e-13}\, T_4^{-0.7}\, \si{cm^3.s^{-1}}\) is
the case~B recombination coefficient and \(T_4 = T/\SI{e4}{K}\).
The advective flux of hydrogen nuclei through the shock is
\begin{equation}
  \label{eq:shell-advective-flux}
  \mathcal{A} = n v_\infty \ ,
\end{equation}
and the flux of hydrogen-ionizing photons
(\(h \nu > \SI{13.6}{eV}\)) incident on the inner edge of the shell is
\begin{equation}
  \label{eq:shell-ionizing-flux}
  \mathcal{F} = \frac{S} {4 \pi R_0^2} \ , 
\end{equation}
where \(S\) is the ionizing photon luminosity of the star.  Any shell
with \(\mathcal{R} + \mathcal{A} > \mathcal{F}\) cannot be entirely
photoionized by the star, and so must have trapped the ionization
front.

The ratio of advective particle flux to ionizing flux is, from
equations~\eqref{eq:Rstar}, \eqref{eq:shell-advective-flux},
\eqref{eq:shell-ionizing-flux},
\begin{equation}
  \label{eq:advective-over-ionizing-flux}
  \frac{\mathcal{A}}{\mathcal{F}} = \num{5.86e-5} \frac{x^2 L_4}{v_{10} S_{49}} \ , 
\end{equation}
where
\begin{equation*}
  S_{49} = S / \bigl( \SI{e49}{s^{-1}} \bigr) \ .
\end{equation*}
Numerical values of \(S_{49}\) for our three example stars are given
in Table~\ref{tab:stars}, taken from Figure~4 of
\citet{Sternberg:2003a}.  Since \(\mathcal{A} \ll \mathcal{F}\) in
nearly all cases, for clarity of exposition we ignore \(\mathcal{A}\)
in the following discussion, although it is included in quantitative
calculations.  The column density of the shocked shell can be found,
for example, from equations~(10) and~(12) of \citet{Wilkin:1996a} in
the limit \(v_\infty/V \to 0\) (Wilkin's parameter \(\alpha\)) and
\(\theta \to 0\).  This yields
\begin{equation}
  \label{eq:shocked-shell-column}
  n\shell h\shell = \tfrac34 n R_0 \ .
\end{equation}
Assuming strong cooling behind the shock
(\S~\ref{sec:radi-cool-lengths}), the shell density is
\begin{equation}
  \label{eq:isothermal-shell-density}
  n\shell = \mathcal{M}_0^2 n \,
\end{equation}
where
\(\mathcal{M}_0 = v_\infty / \sound\) is the isothermal Mach number of the
external stream.\footnote{%
  \label{fn:temperature-dependence}
  The sound speed depends on the temperature and hydrogen and helium
  ionization fractions, \(y\) and \(y_{\text{He}}\) as
  \(\sound^2 = (1 + y + z_{\text{He}} y_{\text{He}}) (k T /
  \bar{m})\), where \(z_{\text{He}}\) is the helium nucleon abundance
  by number relative to hydrogen and
  \(k = \SI{1.3806503e-16}{erg.K^{-1}}\) is Boltzmann's constant.  We
  assume \(y = 1\), \(y_{\text{He}} = 0.5\), \(z_{\text{He}} = 0.09\),
  so that \(\sound = \num{11.4}\, T_4^{1/2}\, \si{km.s^{-1}}\). } %
Putting these together with equations~\eqref{eq:Rstar} and
~\eqref{eq:tau-star}, one finds that \(\mathcal{R} > \mathcal{F}\)
implies
\begin{equation}
  \label{eq:ifront-trap-x-cubed-taustar}
  x^3 \tau_* > \frac{4 S c \sound^2 \bar{m}^2 \kappa}{3 \alpha L} \ .
\end{equation}
From equation~\eqref{eq:rad-full-x}, it can be seen that \(x\) depends
on the external stream parameters, \(n\), \(v_\infty\) only via
\(\tau_*\), and so equation~\eqref{eq:ifront-trap-x-cubed-taustar} is a
condition for \(\tau_*\), which, by using
equation~\eqref{eq:taustar-typical}, becomes a condition on
\(n / v_{10}^2\).  In the radiation bow shock case,
\(x = (1 + \eta\wind)^{1/2}\), and the condition can be written:
\begin{equation}
  \label{eq:ifront-trap-density-RBS}
  \text{RBS:}\quad
  \frac{n}{v_{10}^2} > \num{2.65e8} \, \frac{S_{49}^2 T_4^{3.4}}{L_4^3 (1 + \eta\wind)^3} \ .
\end{equation}
In the radiation bow wave case, \(x = 2\tau_*\), and the condition can be
written:
\begin{equation}
  \label{eq:ifront-trap-taustar-RBW}
  \text{RBW:}\quad
  \frac{n}{v_{10}^2} > \num{5.36e4} \, \frac{S_{49}^{1/2} T_4^{0.85} }{\kappa_{600}^{3/2} L_4^{3/2}} \ . 
\end{equation}
In the wind bow shock case, the result is the same as
equation~\eqref{eq:ifront-trap-density-RBS}, but changing the factor
\((1 + \eta\wind)^3\) to \(\eta\wind^3\).  For the example hot stars in
Table~\ref{tab:stars}, and assuming \(\kappa_{600} = 1\),
\(T_4 = 0.8\), the resulting density threshold is
\(n > (\text{\numrange{1000}{5000}})\, v_{10}^2\), depending only
weakly on the stellar parameters, which is shown by the red lines in
Figure~\ref{fig:zones-v-n-plane}.  For the \SI{10}{M_\odot} star this is
in the radiation bow wave regime, whereas for the higher mass stars it
is in the radiation bow shock regime.  When the external stream is
denser than this, then the outer parts of the shocked shell may be
neutral instead of ionized, giving rise to a cometary compact \hii{}
region \citep{Mac-Low:1991a, Arthur:2006a}.  This is only necessarily
true, however, when the star is isolated.  If the star is in a cluster
environment, then the contribution of other nearby massive stars to
the ionizing radiation field must be considered.

Quite different results are obtained for a B-type supergiant star (see
Tab.~\ref{tab:stars} and Fig.~\ref{fig:B-supergiant}), which has a
similar bolometric luminosity and wind strength to the \SI{40}{M_\odot}
main-sequence star, but a hundred times lower ionizing luminosity.
This results in a far lower threshold for trapping the ionization
front of \(n > 40 v_{10}^2\).  The advective flux, \(\mathcal{A}\), is
relatively stronger for this star than for the main-sequence stars, but
even for \(v_{10} < 2\), where the effect is strongest, the change is
only of order the thickness of the dark red line in
Figure~\ref{fig:B-supergiant}.

In principle, when the ionization front trapping occurs in the bow
wave regime, then the curves for \(R_0\) will be modified in the
region above the red line because all of the ionizing radiation is
trapped in the shell due to gas opacity, which is not included in
equation~\eqref{eq:tau-thin}.  However, this only happens for our
\SI{10}{M_\odot} star, which has a relatively soft spectrum.
Table~\ref{tab:stars} gives the peak wavelength of the stellar
spectrum for this star as \(\lambda_{\text{eff}} = \SI{0.115}{\um}\), which
is significantly larger than the hydrogen ionization threshold at
\SI{0.0912}{\um}, meaning that only a small fraction of the total
stellar luminosity is in the EUV band and affected by the gas opacity.
The effect on \(R_0\) is therefore small.  For the higher mass stars,
\(\lambda_{\text{eff}} < \SI{0.0912}{\um}\), so the majority of the
luminosity is in the EUV band, but in these cases the ionization front
trapping occurs well inside the radiation bow shock zone, where the
dust optical depth is already sufficient to trap all of the radiative
momentum.

\subsection{Efficiency of radiative cooling}
\label{sec:radi-cool-lengths}
In this section, we calculate whether the radiative cooling is
sufficiently rapid behind the bow shock to allow the formation of a
thin, dense shell.  In general, cooling is least efficient at low
densities, so we will assume that the wind bow shock regime applies
unless otherwise specified. We label quantities just outside the shock
by the subscript ``0'', quantities just inside the shock (after
thermalization, but before any radiative cooling) by the subscript
``1'', and quantities after the gas has cooled back to the
photoionization equilibrium temperature by the subscript ``2''.
Assuming a ratio of specific heats, \(\gamma = 5/3\), the relation between
the pre-shock and immediate post-shock quantities
\citep{ZelDovich:1967} is
\begin{align}
  \label{eq:shock-n-jump}
  \frac{n_1}{n_0} &= \frac{4 \M_0^2} {\M_0^2 + 3} \\
  \label{eq:shock-T-jump}
  \frac{T_1}{T_0} &= \tfrac1{16} \bigl( 5\M_0^2 - 1 \bigr) \bigl( 1 + 3/\M_0^2 \bigr) \\
  \label{eq:shock-v-jump}
  \frac{v_1}{v_0} &= \left(\frac{n_1}{n_0}\right)^{-1} \ ,
\end{align}
where \(\M_0 = v_0 / \sound\).  The cooling length of the post-shock
gas can be written as
\newcommand\cool{\ensuremath{_{\text{cool}}}}
\begin{equation}
  \label{eq:dcool}
  d\cool = \frac{3 P_1 v_1} { 2 \bigl(  \mathcal{L}_1 - \mathcal{G}_1 \bigr) }\ ,  
\end{equation}
where \(P_1\) is the thermal pressure and \(\mathcal{L}_1\),
\(\mathcal{G}_1\) are the volumetric radiative cooling and heating
rates.  For fully photoionized gas, we have
\(P_1 \approx 2 n_1 k T_1\), \(\mathcal{L}_1 = n_1^2 \Lambda(T_1)\), and
\(\mathcal{G}_1 = n_1^2 \Gamma(T_1)\), where \(\Lambda(T)\) is the cooling
coefficient, which is dominated by metal emission lines that are
excited by electron collisions, and \(\Gamma(T)\) is the heating
coefficient, which is dominated by hydrogen photo-electrons
\citep{Osterbrock:2006a}. The cooling coefficient has a maximum around
\SI{e5}{K}, and for typical ISM abundances can be approximated as
follows:
\begin{align}
  \label{eq:cooling-coefficient}
  \Lambda_{\text{warm}} &= \num{3.3e-24} \, T_4^{2.3} \, \si{erg.cm^{3}.s^{-1}}\\
  \Lambda_{\text{hot}} &= \num{e-20} \, T_4^{-1}\, \si{erg.cm^{3}.s^{-1}} \\
  \Lambda &= \left( \Lambda_{\text{warm}}^{-k} +  \Lambda_{\text{hot}}^{-k} \right)^{-1/k}
      \quad \text{with} \quad k = 3 \ ,
\end{align}
which is valid in the range \(0.7 < T_4 < 1000\).  We approximate the heating coefficient as
\begin{equation}
  \label{eq:heating-coefficient}
  \Gamma = \num{1.77e-24} \, T_4^{-1/2} \, \si{erg.cm^{3}.s^{-1}} \ ,
\end{equation}
where the coefficient is chosen so as to give \(\Gamma = \Lambda\) at
an equilibrium temperature of \(T_4 = 0.8\).

In Figure~\ref{fig:zones-v-n-plane} we show curves calculated from
equations~\eqref{eq:shock-n-jump} to~\eqref{eq:heating-coefficient},
corresponding to \(d\cool = R_0\) (thick blue line) and
\(d\cool = h_0\) (thin blue line), where \(h_0\) is the shell
thickness in the limit of instantaneous cooling.  In this context,
\(n_0 = n\) and \(n_2 = n\shell\), so that \(h_0\) follows from
equations~\eqref{eq:shocked-shell-column}
and~\eqref{eq:isothermal-shell-density} as
\begin{equation}
  \label{eq:strong-cooling-h0}
  h_0 = \tfrac34 \M_0^{-2} R_0 \ .
\end{equation}
The bends in the curves at \(v \approx \SI{50}{km.s^{-1}}\) are due to the
maximum in the cooling coefficient \(\Lambda(T)\) around
\(\SI{e5}{K}\).  For bow shells with outer stream densities above the thin
blue line, radiative cooling is so efficient that the bow shock can be
considered isothermal, and so the shell is dense and thin (at least,
in the apex region).  It can be seen that the ionization front
trapping always occurs at densities larger than this, which justifies
the use of equation~\eqref{eq:isothermal-shell-density} in the
previous section.  For bow shells with outer stream densities below the
thick blue line, cooling is unimportant and the bow shock can be
considered non-radiative.  In this case the shell is thicker than in
the radiative case,
\(h\shell/R_0 \approx \text{\numrange{0.2}{0.3}}\).  This value can be found
from equation~\eqref{eq:shocked-shell-column} by substituting
\(n = n_0\) and \(n\shell \approx 1.1 n_1\), then using
equation~\eqref{eq:shock-n-jump}. The factor \num{1.1} comes from
consideration of the slight increase in density between the shock and
the contact/tangential discontinuity.  For bow shells with outer stream
densities between the two blue lines, cooling does occur, albeit
inefficiently, so that the shell thickness \(h\shell\) is set by
\(d\cool\) rather than \(h_0\).

\subsection{Axial structure of bow shells}
\label{sec:axial-structure-bow}

\begin{figure}
  \centering
  \includegraphics[width=\linewidth]{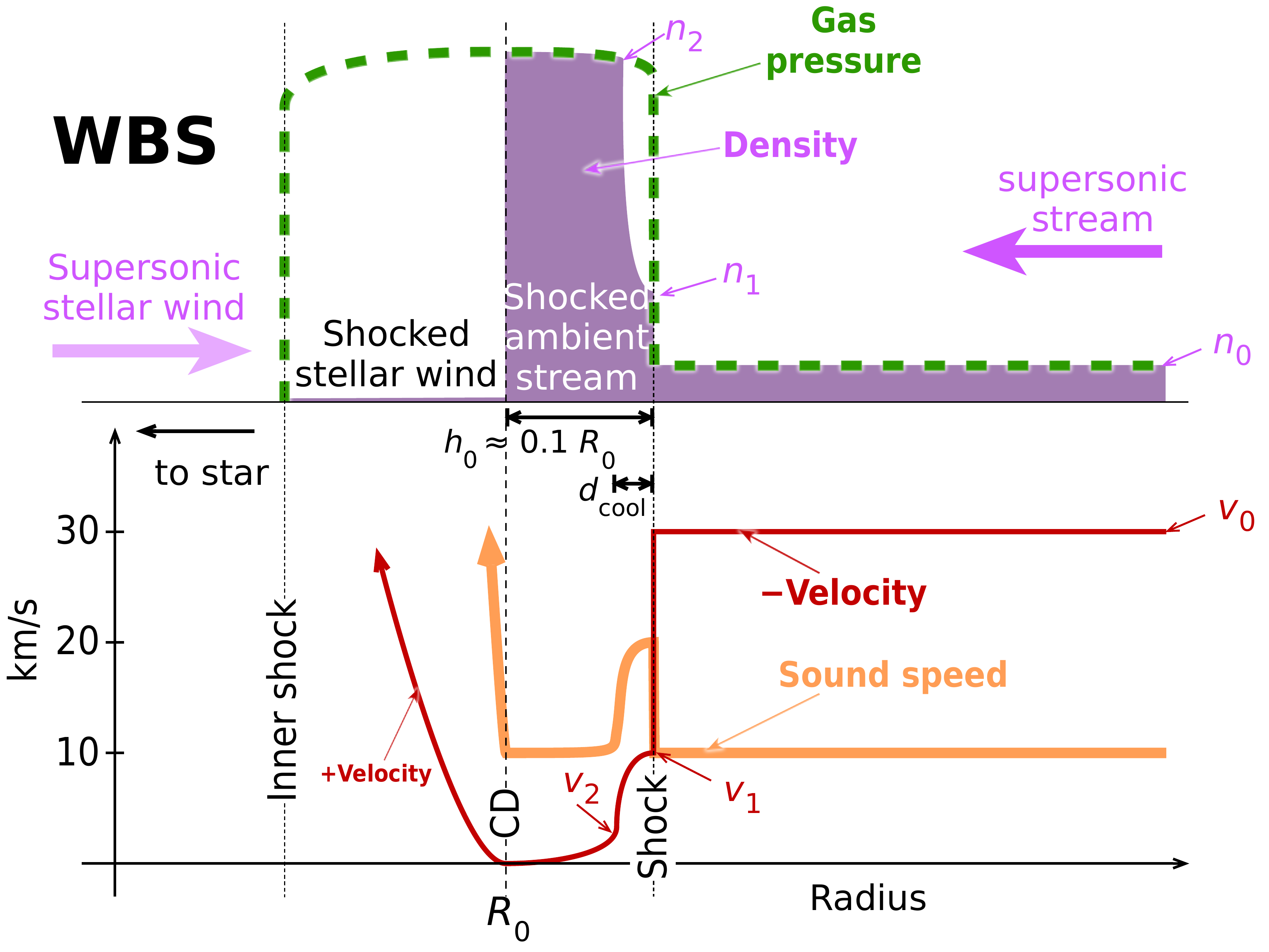}
  \caption{Profiles of physical variables along the axis of the bow
    shell for a wind-supported bow shock.  The top panel shows the gas
    density (purple) and the gas thermal pressure (green dashed line)
    as a function of radius from the star. The bottom panel shows the
    absolute value of the gas velocity (red line) and sound speed
    (orange line).  The stellar wind comes from the left with positive
    velocity, while the ambient stream comes from the right. All
    velocities are in the frame of reference of the star, which is
    located off the left-hand edge of the figure.  The case shown is
    for a stream velocity of \SI{30}{km.s^{-1}} and an equilibrium
    ionized sound speed of \SI{10}{km.s^{-1}}, with a post-shock
    cooling length \(d\cool\) that is smaller than the shell thickness
    \(h_0\). The different flow regions are described in the text.}
  \label{fig:axial-structure-wbs}
\end{figure}
\begin{figure}
  \centering
  \includegraphics[width=\linewidth]{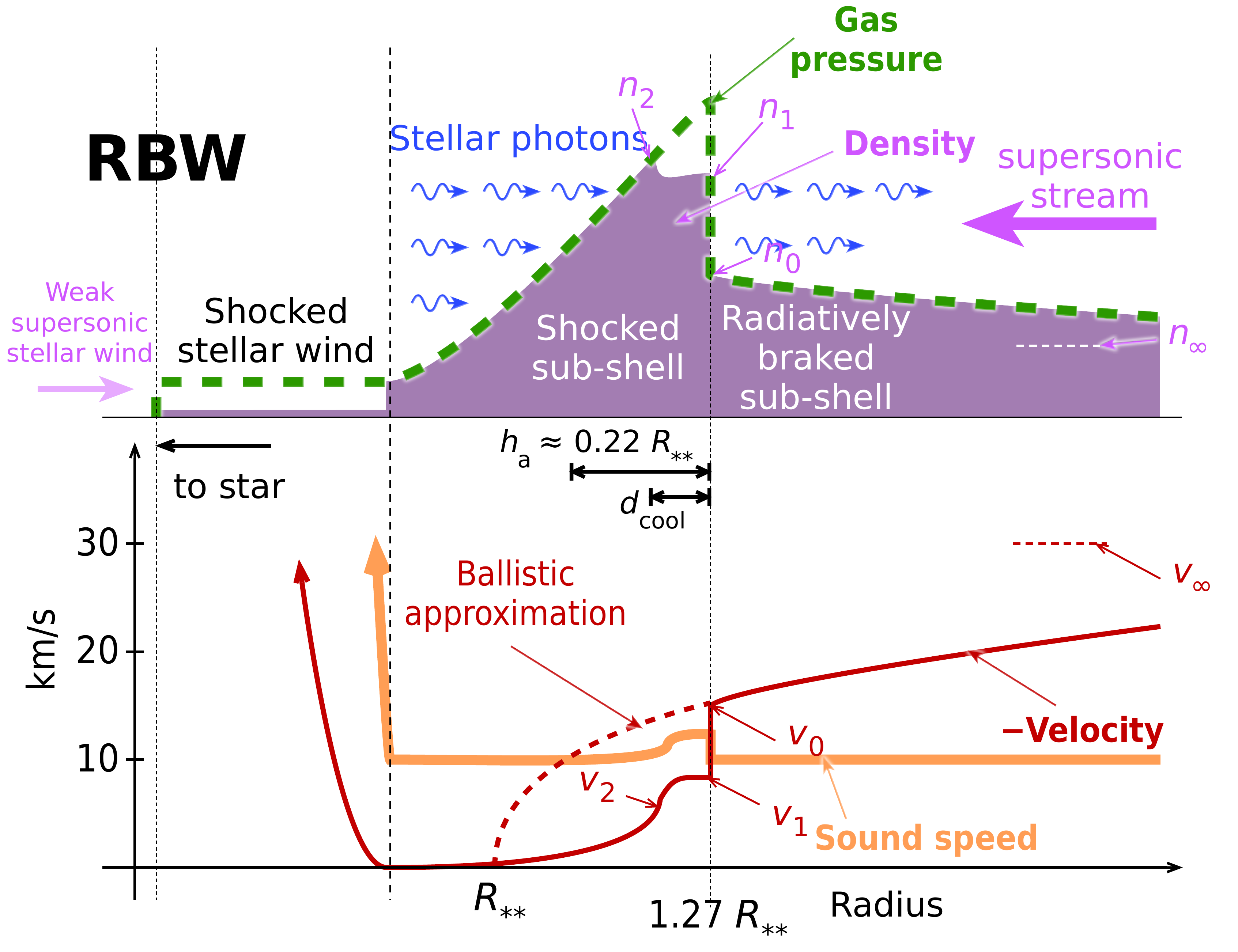}
  \caption{As Fig.~\ref{fig:axial-structure-wbs} but for a
    radiation-supported bow wave.  The heavy dashed red line in the
    lower panel shows the ballistic approximation to the velocity
    profile, which would be followed in the case of a highly
    supersonic incident stream.  The ambient stream is gradually
    decelerated by absorption of stellar radiation, leading to a lower
    velocity and higher density just before the shock, \(v_0\) and
    \(n_0\), as compared to the values at large distances,
    \(v_\infty\) and \(n_\infty\), which are shown by dashed horizontal lines.}
  \label{fig:axial-structure-rbw}
\end{figure}
\begin{figure}
  \centering
  \includegraphics[width=\linewidth]{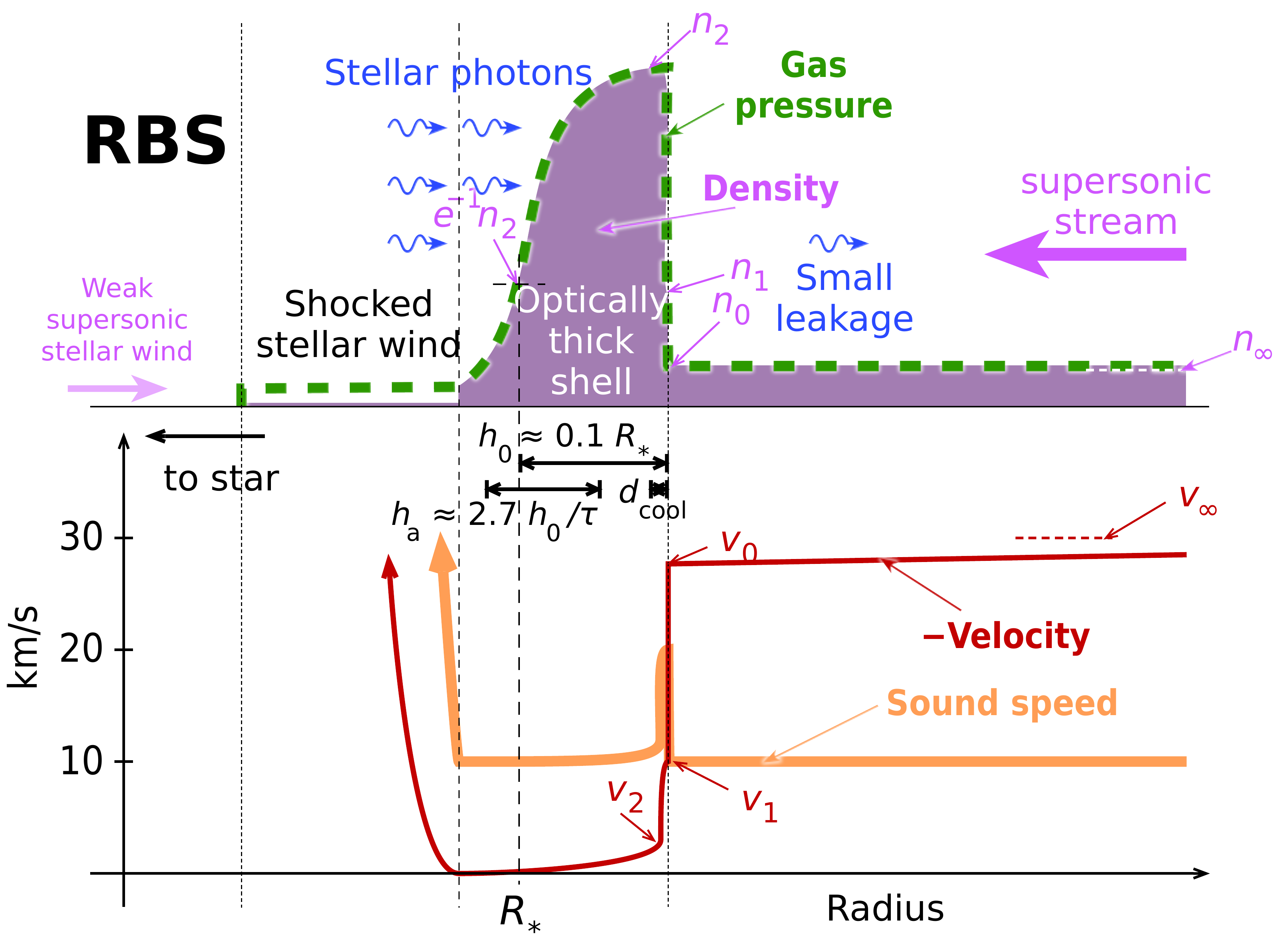}
  \caption{As Fig.~\ref{fig:axial-structure-wbs} but for a
    radiation-supported bow shock.  A smaller cooling length is assumed
    than in the previous figures since the RBS regime occurs at higher
    ambient densities.}
  \label{fig:axial-structure-rbs}
\end{figure}

In the shell of a wind-supported bow shock, the gas density is highest
at the contact discontinuity, but this is no longer true for
radiation-supported bow shells, as illustrated in
Figures~\ref{fig:axial-structure-wbs} to
\ref{fig:axial-structure-rbs}, where we schematically show the shell
structure for the three cases: WBS, RBW, and RBS.  In all cases, we
assume a far-field mach number of \(\M_\infty = 3\) for the incident stream
of density \(n_\infty\).  For the WBS case
(Fig.~\ref{fig:axial-structure-wbs}), the stream is unchanged until it
reaches the shock, so \(\M_0 = \M_\infty\) and \(n_0 = n_\infty\).  The density
increases to \(n_1 = 3 n_0\) in the shock
(eq.~[\ref{eq:shock-n-jump}]), and then to \(n_2 = 9 n_0\)
(eq.~[\ref{eq:isothermal-shell-density}]) after it has cooled back
down to the equilibrium photoionized temperature.  In the figure, we
show the case where \(d\cool < h_0\), so that most of the shell is at
roughly constant density.  Note that although the mass flux along the
axis, \(n v\), is approximately conserved in most of the flow, this is
no longer true close to the contact discontinuity, since the
streamlines bend away from the axis due to lateral pressure gradients.
This is what allows \(v\) to tend to zero, while \(n\) remains
constant.  The thermal pressure in the shocked stellar wind is equal
to the shell pressure, but its density is much smaller due to a
temperature that is higher by a factor of \(\approx 3100 V_3^2\),
(eq.~[\ref{eq:shock-T-jump}], assuming inefficient cooling).

For the optically thin RBW case (Fig.~\ref{fig:axial-structure-rbw}),
equations~\eqref{eq:rad-accel} to~\eqref{eq:rad:R0} can be combined to
give the radial dependence of the stream velocity as it approaches the
star:
\begin{equation}
  \label{eq:v-RBW-ballistic}
  v(R) = \left( 1 - \frac{R_{**}}{R} \right)^{1/2} v_\infty \ .
\end{equation}
This ballistic approximation (shown by the dashed red line in
Fig.~\ref{fig:axial-structure-rbw}) predicts that the velocity smoothly
decreases to zero at \(R = R_{**}\).  However, although this is valid
in the limit of high Mach number, it ignores the effects of gas
pressure and so becomes invalid once the stream velocity falls close
to the sound speed.  The situation bears similarities to the flow in a
supersonic diffuser, such as the inlet of a ramjet or other supersonic
aircraft engine \citep{Seddon:1999a}, which slows supersonic airflow
down to subsonic speeds before combustion.  Exactly the opposite flow
configuration is present in a rocket nozzle \citep{Courant:1948a},
where the flow can smoothly pass from subsonic to supersonic
velocities at the throat of the nozzle.  An analogy between an
isothermal stellar wind and a rocket nozzle is developed in detail in
\S~3.5 of \citet{Lamers:1999b}.  However, in the reverse case of a
supersonic diffuser, a smooth transition from supersonic to subsonic
flow is not possible \citep{Morawetz:1956a}.  Instead, a normal shock
wave develops, which decelerates the supersonically entering flow,
allowing it to exit subsonically \citep{Embid:1984a, Hafez:1999a}.
The supersonic inlet/bow wave analogy is not so exact as the rocket
nozzle/stellar wind analogy due to the multi-dimensional nature of the
bow shell, which means that lateral flows of gas away from the apex
become important when the velocity is subsonic.\footnote{%
  In the aeronautical case, the inlet is usually designed so that
  multiple oblique shocks partially decelerate the flow before it
  passes through the normal shock.  This is the analogy of the initial
  gradual radiative deceleration in the bow wave.  } %
Nonetheless, we expect that a normal shock must be present in the
flow, although there does not seem to be any simple argument for
predicting the exact Mach number where it will occur.  In
Figure.~\ref{fig:axial-structure-rbw} we assume that the shock occurs at
a Mach number \(\M_0 = 1.5\), but multidimensional numerical
simulations are necessary to test this supposition.

The shell in the RBW case therefore consists of two parts: an outer
radiatively braked sub-shell, and an inner shocked sub-shell.  In the
outer sub-shell, the density gradually increases inwards as the stream
is decelerated by the absorbed photons, reaching
\begin{equation}
  \label{eq:outer-shell-density}
  n_0 = \frac{\M_\infty}{\M_0} n_\infty
\end{equation}
just outside the shock (\(n_0 = 2 n_\infty\) for the case illustrated).  In
the inner sub-shell, on the other hand, the pressure must increase
outwards, since it is subsonic and therefore in approximate
hydrostatic equilibrium with an outward-pointing effective gravity
from the radiation force.  The shell thickness will be set by the
hydrostatic scale height, \(h_a = \sound^2 / a\rad\), which by
eqs.~(\ref{eq:rad-accel}, \ref{eq:rad:R0}) is given by
\begin{equation}
  \label{eq:scale-height}
  h_a = \frac{2 R_{**}}{\M_\infty^2} \ .
\end{equation}
If the cooling is efficient (\(d\cool < h_a\)), then most of the
sub-shell is isothermal and the density will fall off exponentially
towards the star.  The contact discontinuity with the stellar wind
will form at the point where the shell pressure has fallen by a factor
of order \(\tau / \eta\wind\), but this has no influence on the bulk of the
shell, which is pressurized by radiation, not wind.  If, as we
suspect, \(\M_0\) depends only weakly, if at all, on \(\M_\infty\), then
eqs.~(\ref{eq:outer-shell-density}, \ref{eq:scale-height}) imply that
the inner shocked sub-shell represents a fraction
\(\approx \M_\infty^{-1}\) of the total shell optical depth, which means that the
outer supersonic sub-shell dominates for highly supersonic stream
velocities.

For the optically thick RBS case (Fig.~\ref{fig:axial-structure-rbs}),
the shell density profile is no longer simply exponential since the
pressure scale height now increases as one moves away from the star,
due to the extinction-induced decline in radiation flux. Assuming
single scattering and plane geometry, the resultant hydrostatic
density profile is doubly exponential:
\begin{equation}
  \label{eq:rbs-density-profile}
  n(R) = n_2 e^{-e^{-x}} \text{\quad where\quad } x = n_2 \bar{m} \kappa (R - R_*) \ . 
\end{equation}
The shell density increases from \(e^{-1} n_2\) at \(R_*\) to saturate
at \(n_2\) for \(R > R_* + h_a\), where \(h_a\) is now the pressure
scale height at the inner edge of the shell.  From
equations~(\ref{eq:Rstar}, \ref{eq:rad:R0}, \ref{eq:tau-thin},
\ref{eq:strong-cooling-h0}, \ref{eq:scale-height}), and taking
\(R_0 \approx R_*\), we find
\begin{equation}
  \label{eq:ha-versus-h0}
  h_a \approx \frac{4 h_0}{3 \tau} \ .
\end{equation}
We show results for a total shell optical depth \(\tau = 3\), for which
\(h_a\) is a significant fraction of \(h_0\).  For much larger optical
depths, \(h_0 \gg h_a\) so that the constant density portion of the
shell will dominate the total column.  A small fraction \(e^{-\tau}\) of
the photons will penetrate the shell and be available to decelerate
the supersonic stream, as in the RBW case, yielding
\(v_0 = (1 - \tau e^{-\tau})\, v_\infty\).  In the illustrated case of
\(\tau = 3\), the resultant speed reduction before the shock is only 8\%.
As mentioned in \S~\ref{sec:effects-gravity}, radiative acceleration
becomes less important than gravity in the outer regions of an opaque
shell.  However, in order for this to have a significant effect on the
shell, the gravitational scale height would have to be be less than
the shell thickness: \(\Gamma\Edd h_a < h_0\).  From
equation~\eqref{eq:ha-versus-h0} this translates to
\(\tau > \frac43 \Gamma\Edd\), where dust-opacity Eddington factors are
\(\Gamma\Edd \sim 100\) for OB stars.  Such large optical depths would
require extremely high ambient densities of
\(n > \num{2e7} L_4 \,v_{10}^2 \left( \SI{10}{M_\odot}/M \right)^2 \,
\si{cm^{-3}} \), where we have used equations~(\ref{eq:tau-star},
\ref{eq:eddington-factor}) and that \(\tau \approx 2\tau_*\) in the RBS limit.

\section{Discussion}
\label{sec:discussion}

The stellar bow shells modeled in this paper can be seen as either due to
the motion of a star through the interstellar medium, or due to the
motion of the interstellar medium that flows past the star.  The
first case corresponds to runaway stars \citep{Blaauw:1961a}, which
have been ejected from a binary system or stellar cluster
\citep{Hoogerwerf:2001a}, while the second case can be due to
photoevaporation and champagne flows in \hii{} regions
\citep{Tenorio-Tagle:1979a, Shu:2002a, Henney:2005a}, or to general
turbulent flows in the Galaxy
\citep[e.g.,][]{Ballesteros-Paredes:1999a}.  In this section, we
consider the stream densities and velocities expected in each of these
scenarios, and compare them with our predictions for the type of bow shells
that should result.

Although some runaway stars have velocities exceeding
\SI{100}{km.s^{-1}}, most are moving considerably slower, with a
median peculiar velocity of about \SI{30}{km.s^{-1}}
\citep{Tetzlaff:2011a}.  The higher velocity runaways are likely
produced by dynamical interactions in the center of young clusters
\citep{Gualandris:2004a}, whereas the supernova-induced dissolution of
binary systems is predicted to primarily produce ``walkaways'' with
even slower velocities of order \SI{10}{km.s^{-1}}
\citep{Renzo:2018a}.  Environmental flows are also expected to be of
order \SI{10}{km.s^{-1}}, which is a characteristic velocity
dispersion for warm neutral gas in the inner Galaxy
\citep{Marasco:2017a} and also a typical expansion velocity for
\ion{H}{i} shells \citep{Ehlerova:2005a}.  Internal velocity
dispersions within \hii{} regions are subsonic
(\SIrange{5}{10}{km.s^{-1}}) for small regions, such as the Orion
Nebula, containing one or a few ionizing stars
\citep{Arthur:2016a}. This increases slowly for more luminous regions
as \(\sigma \propto L^{1/4}\) \citep{Bordalo:2011a}, so giant star forming
complexes such as Carina (100 times more luminous than Orion) show
velocity dispersions of order \SI{20}{km.s^{-1}}
\citep{Damiani:2016a}.  Higher velocities of
\SIrange{30}{40}{km.s^{-1}} are reached in divergent photoevaporation
flows \citep{Dyson:1968a}, but this is achieved at expense of a lower
density.  Given all the above, we expect there to be many more bow shells
with relative stream velocity \(v_\infty \sim \SI{20}{km.s^{-1}}\) than bow shells
with \(v_\infty \sim \SI{100}{km.s^{-1}}\).

Turning now to the ambient density, if bow shock stars were randomly
sampling the volume of the Galactic disk, then we would expect the
average density to be \(< \SI{1}{cm^{-3}}\).  The dominant gas phase
\citep{Ferriere:2001a} near the Galactic mid-plane is the Warm Neutral
Medium with a volume filling fraction of 45\% \citep[Fig.~11
of][]{Kalberla:2009a} and an average density of \SI{0.9}{cm^{-3}} at
the Solar circle \citep[\S~4 of][]{Kalberla:2008a}.  Significant
fractions of the volume (\(\approx 20\%\) each) are also occupied by the
Warm Ionized Medium and Hot Ionized Medium, which have even lower
densities (\(\approx \SI{0.3}{cm^{-3}}\) and \SI{e-3}{cm^{-3}},
respectively) and which increasingly dominate the volume for heights
\(z > \SI{500}{pc}\) above the plane.  The denser Cold Neutral Medium
(\(n \approx \text{\SIrange{10}{100}{cm^{-3}}}\)) and Molecular Clouds
(\(n > \SI{1000}{cm^{-3}}\)) occupy much smaller volumes
(\(\approx 10\%\) and \(< 1\%\), respectively).  However, observations of
stellar bow shells are clearly biased towards these higher densities.

In part, this bias is due to bow shells being easier to detect in denser
environments.  Depending on the emission mechanism, the shell
luminosity will be proportional to the column density
(\(\propto n R_0\)) or volume emission measure
(\(\propto n^2 R_0^3\)), which are both increasing functions of \(n\) since,
from \S~\ref{sec:three-bow-regimes}, the bow shell size falls relatively
slowly as \(R_0 \propto n^{-1/2}\) in the WBS and RBS cases, and is
independent of \(n\) in the RBW case.  Another contribution to the
high-density bias is simply that none but the highest velocity
high-mass stars can move far during their lifetime from the molecular
clouds where they were born.  Even a runaway star with
\(v = \SI{30}{km.s^{-1}}\), will travel \(< \SI{120}{pc}\) during an
O-star lifetime of \(< \SI{4}{Myr}\), which is of the same order as
the sizes of Giant Molecular Clouds.  Many observed stellar bow shells are
found in high mass star clusters associated with large \hii{} regions
\citep{Povich:2008a, Sexton:2015b}.  The ionized gas density in
Carina, for example, has an average value of
\(\approx \SI{100}{cm^{-3}}\) \citep{Oberst:2011a, Damiani:2016a}, although
with peaks \(> \SI{3000}{cm^{-3}}\) in photoevaporation flows from
embedded molecular globules \citep{Smith:2004a}.  In more compact
\hii{} regions, such as the Orion Nebula, ionized densities up to
\SI{e4}{cm^{-3}} are found on scales of about \(\SI{0.1}{pc}\)
\citep{Weilbacher:2015a, ODell:2017b}.

In \S~\ref{sec:depend-stell-type} we found that the bow shell is
wind-supported when \(\tau_*^2 < \eta\wind\), where \(\tau_*\) is a fiducial
optical depth (eq.~[\ref{eq:taustar-typical}]) and \(\eta\wind\) is the
radiative momentum efficiency of the stellar wind
(eq.~[\ref{eq:wind-eta-typical}]).  This translates into a maximum
stream density for wind-supported bow shells of
\newcommand\crit{\ensuremath{_{\text{max}}}}
\newcommand\vink{\ensuremath{_{\text{\tiny Vink}}}}
\begin{equation}
  \label{eq:ncrit}
  n\crit \approx \left(\SI{30}{cm^{-3}}\right)
  \left(  \frac{v} {\SI{10}{km.s^{-1}}}\right)^2
  \left( \frac{\eta\wind}{\eta\vink} \right)
  \left( \frac{\SI{600}{cm^2.g^{-1}}}{\kappa} \right)
  \ ,
\end{equation}
which is largely insensitive to spectral type between early-B and
early-O stars.  In this equation, we have expressed the wind
efficiency in terms of the Table~\ref{tab:stars} values,
\(\eta\vink\), which are calculated from the \citet{Vink:2000a} recipe,
and the UV dust opacity in terms of the standard ISM value adopted in
\S~\ref{sec:depend-stell-type}.

Therefore, runaway stars moving through the diffuse ISM with
\(n \sim \SI{1}{cm^{-3}}\) and \(v \sim \SI{30}{km.s^{-1}}\) are safely in
the WBS regime, having \(n \approx \num{0.004}\, n\crit\) for the default
wind and dust parameters.  On the other hand, slow-moving stars in
\hii{} regions of moderate density (\(n \sim \SI{100}{cm^{-3}}\)) can
easily have \(n > n\crit\), meaning that their bow shells will be radiation
supported.

\begin{figure}
  \includegraphics[width=\linewidth]{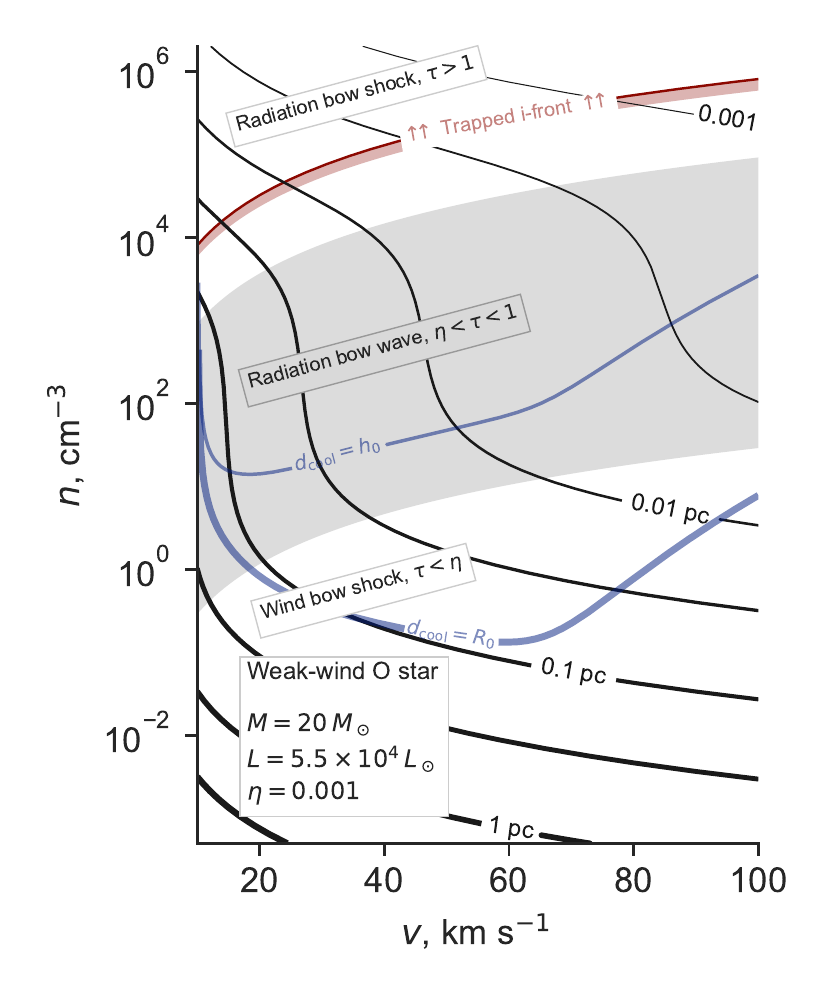}
  \caption{As Fig.~\ref{fig:zones-v-n-plane}, but for a weak-wind
    O~star, similar to the sample observed by \citet{Martins:2005b}.
    For this star, radiation-supported bow waves (gray shading) occur
    over a much larger region of parameter space than for a
    normal-wind star of the same spectral type
    (see Fig.~\ref{fig:zones-v-n-plane}b).}
  \label{fig:O-weak-wind}
\end{figure}

For cases where the stellar wind or dust properties differ from our
assumed values, then these conclusions may change.  For instance,
there are a growing number of stars for which very low mass-loss rates
have been diagnosed from UV, optical or infrared line profiles
\citep{Martins:2005b, Marcolino:2009a, Najarro:2011a, Martins:2012a,
  Shenar:2017a, Smith:2017b}.  Most of these stars are late-type
O~dwarfs, for which the \citet{Vink:2000a} prescription predicts a
mass-loss rate of \num{3e-8} to \SI{e-7}{M_\odot.yr^{-1}}, whereas the
observationally derived values are in the range \num{e-10} to
\SI{3e-9}{M_\odot.yr^{-1}} -- a shortfall of 10 to 1000!  This ``weak wind
problem'' is a far larger discrepancy than can be explained by
clumping effects.  A potential (partial) resolution is to suppose that
internal shocks near the base of the wind, together with thermal
conduction, heat a large fraction of the gas to coronal temperatures,
where it is no longer detectable by means of standard wind diagnostic
lines \citep{Lucy:2012a}.  However, in cases where x-ray diagnostics
have been used to trace the mass loss of this hot component
(\(\mu\)~Col, \citealp{Huenemoerder:2012a}; \(\zeta\)~Oph,
\citealp{Cohen:2014a}), rates that are still an order of magnitude
below the Vink values are found (see also \citealp{Shenar:2017a}).  In
Figure~\ref{fig:O-weak-wind}, we show results for a weak-wind O9
dwarf, which is identical to the \SI{20}{M_\odot} star from
Table~\ref{tab:stars}, except with \(\eta\wind = 0.01 \eta\vink\)
(approximating the measurements from the sample of
\citealp{Martins:2005b}).  For such stars, the boundary between the
wind-supported and radiation-supported regimes shifts to lower ambient
densities, reaching \(n\crit \sim \SI{1}{cm^{-3}}\) for
\(v = \SI{20}{km.s^{-1}}\).  This means that radiation-supported bow shells
become feasible even in the diffuse ISM for this important class of
stars (if winds are really as weak as the UV diagnostics suggest).  It
is therefore important to critically re-evaluate attempts to
\emph{measure} mass loss rates using observations of bow shells
\citep{Gvaramadze:2012a, Kobulnicky:2018a} since such calculations
invariably assume that the WBS regime is universally applicable.  This
issue is addressed in detail in Paper~III.

Variations in the UV dust opacity would also effect our results.
These may be due to variations in the dust/gas ratio or to changes in
the composition or size distribution of the grains.  The UV opacity is
dominated by very small grains (VSG) with radius
\(a \sim \SI{10}{nm}\), which represent a small fraction
(\(\simeq 5\%\)) of the total dust mass in standard grain mixtures
\citep{Desert:1990a}.  This same VSG population dominates the
mid-infrared continuum emission around \SI{24}{\um}, which is where
stellar bow shells are most easily detected \citep{Meyer:2016a,
  Kobulnicky:2016a}. There is some evidence that such grains may be
depleted in the ionized gas of the Orion Nebula \citep{Salgado:2016a},
leading to a decrease in \(\kappa\) by a factor of about six.  By
equation~\eqref{eq:ncrit}, this would increase \(n\crit\) by the same
factor, making radiation-supported bow shells less likely.  On the other
hand, the process of Radiative Torque Disruption is predicted to be
important \citep{Hoang:2018a} for the high radiation fields found in
stellar bow shells, and this would tend to \emph{increase} the VSG abundance
(and hence \(\kappa\)) via the centrifugal destruction of larger grains.
In LMC \hii{} regions, \citet{Stephens:2014b} find observational
evidence that the VSG abundance increases as the radiation field gets
stronger.

Variations in metallicity will also be reflected in \(\kappa\), assuming
that the total dust-gas ratio is roughly proportional to the metal
abundance, \(Z\).  However, the wind strength \(\eta\wind\) also
increases with metallicity (\(\dot{M} V\wind \propto Z^{0.82}\),
\citealp{Vink:2001a}), which will largely cancel out the effect in our
bow shell models.




\section{Summary}
\label{sec:summary}

We have presented a systematic study of the formation of emission
arcs, or bow shells, around luminous stars that move supersonically
relative to their surrounding medium, taking into account both the
stellar wind and radiation pressure of the star.  In this initial
study, we considered the case where gas and grains are perfectly
coupled via collisions, and applied our models to OB stars.  Our
principal results are as follows:
\begin{enumerate}[1.]
\item Three different regimes of interaction are possible, in order of
  increasing optical depth of the bow shell: Wind-supported Bow Shock
  (WBS), Radiation-supported Bow Wave (RBW), and Radiation-supported
  Bow Shock (RBS).
\item For a broad range of stellar types, the WBS regime occurs when
  the ambient density is below a critical density:
  \(n\crit \approx \SI{100}{cm^{-3}}\) for a typical relative velocity of
  \SI{20}{km.s^{-1}}.
\item The critical density increases for faster moving stars, but
  decreases for stars with weak stellar winds.
\item At densities higher than \(n\crit\), B~stars tend to be in the
  RBW regime, and O~stars in the RBS regime.
\item For main sequence OB stars, the bow shell remains fully
  photoionized for ambient densities up to about 100 times
  \(n\crit\). For isolated B~supergiants, on the other hand, the
  ionization front may be trapped by the shell even in the WBS regime.
\item The bow shell in the RBW regime tends to be broad, with the
  density gradually ramping up at the inner and outer edges, whereas
  in the WBS and RBS regimes the shell is thinner with more sharply
  defined edges.
\item Studies that have estimated wind mass-loss rates from
  observations of bow shells need to be re-evaluated in the light of
  possible radiation support. 
\end{enumerate}


\section*{Acknowledgements}
We are grateful for financial support provided by Dirección General de
Asuntos del Personal Académico, Universidad Nacional Autónoma de
México, through grants Programa de Apoyo a Proyectos de Investigación
e Inovación Tecnológica IN112816 and IN107019.  We thank the referee
for a helpful report.

\bibliographystyle{mnras}
\bibliography{bowshocks-biblio}

\bsp	
\label{lastpage}
\end{document}